\documentclass[preprint,showpacs,preprintnumbers,amsmath,amssymb]{revtex4}

\usepackage{graphics}
\usepackage{dcolumn}
\usepackage{bm}

\begin{document}
%
\title  {THE EXOTIC STATISTICS OF LEAPFROGGING SMOKE RINGS }

\author{Antti J. Niemi } 

\email{niemi@teorfys.uu.se}

\affiliation{ Department of Theoretical Physics, 
Uppsala University, P.O. Box 803, S-75108, Uppsala, Sweden }

\homepage{http://www.teorfys.uu.se/people/antti}

\date{\today}

\begin{abstract}
The leapfrogging motion of smoke rings is a three dimensional 
version of the motion that in two dimensions leads 
to exotic exchange statistics. The statistical 
phase factor can be computed using the hydrodynamical Euler equation,
which is a universal law for describing the properties of a
large class of fluids. This suggests that three dimensional exotic 
exchange statistics is a common
property of closed vortex loops in a variety 
of quantum liquids and gases, from helium superfluids to 
Bose-Einstein condensed alkali gases, metallic hydrogen 
in its liquid phases and maybe even nuclear matter in extreme conditions.

\end{abstract}

\maketitle

%
%

Closed vortex loops such as an ordinary smoke ring, are
common in inviscid and low viscosity fluids. They are
among the simplest examples of organized structure formation, 
and increasingly important in a variety of scenarios.
In cosmology, the theory of cosmic string loops 
might explain a myriad of phenomena from baryon number 
asymmetry to galactic structure formation \cite{shell}. 
In strong interaction physics the string that confines quarks 
might also exist in solitude as a closed string that 
describes glueballs \cite{lud}, and the physical 
principles that govern the behaviour of the confining
string can also explain the origin of most of the mass
observed in the Universe \cite{clay}. 
In condensed matter physics, the properties of closed 
vortex loops are under an intense scrutiny in a variety 
of fluids, including quantum superfluids \cite{donn}, atomic 
Bose-Einstein condensed alkali gases \cite{leg}, nematic liquid 
crystals \cite{bow}, and 
the liquid phases of metallic hydrogen \cite{egor}. Finally, the 
dynamics of vortex loops is highly relevant to hydrodynamical
turbulence \cite{fluid}, which is 
widely considered as {\it the} unexplained phenomenon in 
classical physics \cite{clay}. 

The interactions between vortex loops can
lead to stunning phenomena. One of the most 
impressive sights occurs when two 
identical smoke rings leapfrog through each other
as shown in Figure 1.
\begin{figure}[h]
\includegraphics{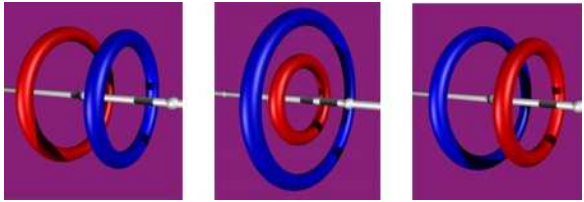}
\caption{The leapfrogging motion of two smoke rings along
their common axis begins with the rear (red) ring 
contracting and accelerating, at the same time 
the front (blue) ring slows down and expands 
its diameter (left picture). Eventually the rear ring catches up with 
the front ring, slides through its center (center picture), 
and emerges ahead of it (right picture). This reverses the role of the 
smoke rings and under ideal conditions
the process repeats itself almost perpetually. 
}
\end{figure}

In the present letter we propose that the leapfrogging motion 
is the three dimensional version
of the motion that in a two dimensional context leads
to exotic exchange statistics \cite{wilc}:

In certain two dimensional physical systems such as very 
thin ($^4$He) superfluid films, fractional quantum Hall systems, 
and high temperature superconductors, identical point particles 
which are subject to the laws of quantum mechanics 
do not need to obey the conventional bosonic and fermionic exchange 
statistics. Instead, under exchange their quantum mechanical 
wave function acquires a phase factor
$(-1)^\Phi$, where the statistical phase $\Phi$ can have
{\it any} value \cite{wilc}. In principle, this phase can be made
visible by interference, when a pointlike particle 
such as a point
vortex in a thin $^4$He superfluid film traverses 
around a closed loop $\Gamma$. The quantum 
mechanical phase $\Phi$ displays
the following two (universal) contributions \cite{haldane}, 
\cite{wilc}
\begin{equation}
\Phi = 2\pi Q ( \rho_0 A  - n \cdot \pi \xi^2 \ \rho_V )   
\label{2d}
\end{equation}
Here $A$ is the area encircled by $\Gamma$, $Q$ is the circulation 
of the vortex which is being traversed
as a multiplet of $h/m$ with $m$ the mass of the background
fluid particles, and $\rho_0$ is the constant London limit 
density of the background fluid. The integer $n$ counts the 
number of vortices with circulation $Q$ that are being encircled 
by $\Gamma$. The parameter $\xi$ is the healing length that 
characterizes the thickness of the vortex core, and $\rho_V$ 
is an averaged deficit in the density of background fluid 
particles inside the core. 

The first term in (\ref{2d}) is a Bohm-Aharonov phase factor 
for transporting an isolated pointlike vortex around $\Gamma$. 
It is insensitive to the presence of additional vortices and does not
contribute to any exotic statistics. The second term in (\ref{2d}) 
relates directly to the presence of vortices that reside in the 
area $A$, and this is the contribution that leads to exotic
statistics. 

The result (\ref{2d}) idealizes a vortex as a circular object
with an effective core radius $\xi$ which acts
as a short distance cut off scale beyond which a
model dependent description of the vortex structure becomes 
relevant. The {\it universal} feature of (\ref{2d}) is
that independently of model dependent details, 
the exotic contribution to statistics measures 
how the presence of vortices influences the number of 
background fluid particles within the area that is being
encircled by $\Gamma$. 

We shall now proceed to derive 
an analogous universal law in the three dimensional context:

In a three dimensional physical world 
the exchange of pointlike particles is 
topologically trivial. The pertinent quantum mechanical 
wave function can only exhibit 
either the Bose-Einstein ($\Phi = 0$) or the Fermi-Dirac ($\Phi = 1$) 
statistics, even though this point of view has been occasionally 
challenged \cite{hald2}. But if instead of pointlike 
particles we consider the exchange of closed three dimensional 
loops and allow for motions such as the leapfrogging
motion of two smoke rings, the exchange topology becomes nontrivial. 
In fact, the motion group of leapfrogging loops contains Artin's braid
group \cite{dahm} which in the two dimensional context
leads to a calculable exotic exchange statistics. This suggests 
that closed three dimensional loops
should also exhibit exotic statistical behaviour. In fact, in
\cite{bala} a scenario has been proposed for computing 
the ensuing statistical phase in the case of closed strings. 
But the approach in \cite{bala} assumes that the strings carry a nontrivial
internal structure. As a consequence the result is a reflection 
of phenomena which reside in different dimensionalities, and
in a hydrodynamical context the approach would lead to a 
conflict with Kelvin's circulation theorem.

Here we show that in fluids, the exchange of structureless 
closed vortex loops leads to computable exotic exchange statistics. 
The ensuing statistical phase can be evaluated directly
from the hydrodynamical Euler equation which is widely 
considered as {\it the} universal equation for describing 
homogeneous and inviscid liquids and gases above microscopic 
distance scales. As a consequence our result suggests
that three dimensional exotic exchange statistics is commonplace,
it can be present whenever closed vortex loops are exchanged.

The Euler equation reads \cite{fluid}
\begin{equation}
\dot { \bf u} + {\bf u} \cdot \nabla {\bf u} \ = \ - \frac{1}{\rho}
\nabla p + \frac{1}{\rho} {\bf F}  
\label{euler}
\end{equation}
where the fluid velocity ${\bf u}$ and the fluid density 
$\rho$ are subject to the continuity equation
\begin{equation}
\dot \rho + \nabla \cdot ( \rho {\bf u})  = 0
\label{cont}
\end{equation}
For us the details of the (conservative) external force ${\bf F}$
in (\ref{euler}) are quite inessential. For simplicity 
we assume that the flow is isentropic so that any external 
force can be combined with the pressure $p$ into 
an enthalpy $V(\rho)$ which only depends on the density $\rho$. If 
$V(\rho) \sim \lambda (\rho^2 - \rho_0^2)^2$ with $\lambda$ large, 
the fluid density is then practically a constant 
$\rho({\bf x}) \approx \rho_0$ and this defines the London limit.

The version of (\ref{euler}) and (\ref{cont}) which is 
relevant for describing the fluid above microscopic scales 
derives from the classical action
\begin{equation}
S = \int d^3\! x dt \left\{ \rho \dot \theta - \theta \nabla \cdot
(\rho {\bf u} ) - \frac{1}{2} \rho {\bf u}^2 - V[\rho] \right\}
\label{eulact1}
\end{equation}
Indeed, a variation of $\theta$ in (\ref{eulact1}) yields 
(\ref{cont}), a variation of $\bf u$ gives
\begin{equation}
{\bf u} \ = \ \nabla \theta \ \ \ \ \ {\rm when} \ \ \ \rho({\bf x}) \not=0
\label{irro}
\end{equation}
and a variation of $\rho$ leads to the Euler equation (\ref{euler}) 
when we use (\ref{irro}). Note that the action (\ref{eulact1}) has the 
standard Hamiltonian form with $\theta$ 
and $\rho$ as canonical conjugates. Consequently it allows, in 
principle, for a quantum mechanical inspection of
the exchange statistics of vortex loops in fluids which are subject
to quantum laws.

Instead of the (universal) Euler equation we could certainly employ
some  more sophisticated model. In a refined
analysis this may even become practical
since among its limitations, (\ref{eulact1}) yields a gapless energy
spectrum. But since the effect we consider has a long distance component
that would not lead to any modification in our main conclusions:
The Euler equation is a universal model for describing the behaviour
of more elaborate quantum fluids at large length scales, and the very fact
that it encodes exotic statistics is a clear indication 
that our observations are similarly universal and have a wide applicability.

A closed vortex loop defines a region in the fluid where
vorticity is nonvanishing, $\nabla \times {\bf u} \not=0$.  
Since (\ref{irro}) states that vorticity vanishes
whenever $\rho({\bf x}) \not= 0$, for consistency the fluid density 
$\rho({\bf x})$ must vanish at the core of the vortex loop. 
For simplicity we imagine the vortex loop as a very thin, 
but not necessarily very short tubular region in which the
vorticity has its support and the fluid density $\rho({\bf x}) 
\approx 0$. The thickness $\xi$ of the tube 
defines a short distance cut off scale 
beyond which (\ref{eulact1}) lacks validity. In the presence 
of closed vortex loops the region where $\rho \not=0$ 
is then multiply connected, the circulation of ${\bf u}$ 
along irreducible curves that encircle the vortex loops 
does not need to vanish, and the variable $\theta$ becomes in general
multivalued. 

We realize the present scenario by describing a thin vortex tube as a 
closed curve ${\bf R}(s)$ where $s$ measures distance along the curve. 
Our consistency requirement for the fluid density states that on
these curves $\rho[{\bf R}(s)] =0$ while outside of the vortices 
the density very rapidly (over a distance $\approx \xi$) 
approaches its constant London limit value 
$\rho({\bf x}) \approx \rho_0$.

In the vicinity of a vortex loop the velocity 
${\bf u}({\bf x})$ is computed from the Biot-Savart law \cite{fluid}
\begin{equation}
{\bf u}({\bf x}) = Q \cdot \nabla \times
\oint \frac{d{\bf s}}{|{\bf x} - {\bf R}(s)|} + \nabla \psi
\label{biot}
\end{equation}
where $Q$ is ($4\pi$ times) the circulation of the 
vortex loop, and $\psi$ is some
(single valued) function in the fluid that we 
include for completeness as
the fluid does not need to be incompressible. 
The vorticity has indeed support only on the curve ${\bf R}(s)$
\begin{equation}
\nabla \times {\bf u} \ = \ Q \cdot
\oint d{\bf R} \ \delta({\bf x} - {\bf R})
\label{vort}
\end{equation}
and outside of the curve we have indeed a potential flow, 
with velocity field ${\bf u}$ given by
\begin{equation}
{\bf u}({\bf x}) \ = \ - Q \cdot \nabla \Omega ({\bf x}) 
+ \nabla \psi
\label{solid}
\end{equation}
where $\Omega({\bf x})$ is the (signed)
solid angle at the point ${\bf x}$ which is subtended 
by some two dimensional surface $\Sigma$ which is bounded 
by ${\bf R}(s)$. The solid angle has a number 
of valuable properties which are 
obtained by inspecting the two-form \cite{courant}
\begin{equation}
\hat \omega({\bf x}_0) \ = \ \frac{1}{2} \epsilon_{ijk}
\frac{ (x^i- x^i_0) dx^j\wedge dx^k }{|{\bf x} - {\bf x}_0|^3}
\label{2form}
\end{equation} 
Its integral 
\begin{equation}
\Omega ({\bf x}_0)
= \int\limits_{\Sigma} \!\! d^2\! \sigma
\epsilon^{ab}  \epsilon_{ijk}
\frac{\partial z^i}{\partial \sigma^a}
\frac{\partial z^j}{\partial \sigma^b}
\frac{x^k-z^k}{|{\bf x} - {\bf z}|^3}
\label{2form}
\end{equation}
over the surface $\Sigma$ coincides with the (signed)
solid angle which is subtended by $\Sigma$ at ${\bf x}_0$.
Since $d\hat \omega = 0$ the solid angle remains 
intact under such local deformations of the surface $\Sigma$ 
that leave the boundary curve ${\bf R}(s)$ intact: The solid angle depends 
only on the boundary curve, and when we move around the 
boundary curve by linking it once the solid angle 
jumps by $\pm 4\pi$.  Consequently in the presence of 
closed vortex loops the variable $\theta$
in (\ref{irro}) indeed becomes multivalued. 

We now consider an (imaginary) periodic adiabatic motion of a closed vortex 
loop ${\bf R}(s)$ in the fluid. We parametrize the motion by 
${\bf R}(s,t)$ where $t$ is the adiabatic time, and since the motion
is periodic we have ${\bf R}(s,0) = {\bf R}(s,T)$ for some $T$. 
The surface ${\bf R}(s,t)$ encloses a toroidal volume $V_T$ in 
the fluid. We first assume that $V_T$ does not intersect or enclose         
any other vortices. In the London limit we then get from (\ref{eulact1}), 
(\ref{irro}), (\ref{solid}) for the quantum mechanical
adiabatic phase \cite{wilc}
\begin{equation}
\Phi = \int \! d^3\! x \! dt \ \rho \dot \theta 
\nonumber
\end{equation}
\begin{equation}
= -\lim_{\Delta t \to 0}
Q \cdot
\int d^3 \! x \int\limits_0^T dt \ 
\rho({\bf x},t) \frac{ \Omega({\bf x},t+\Delta t) -
\Omega({\bf x},t)}{\Delta t}
\label{phas}
\end{equation}
The difference $\Omega(t+\Delta t) - \Omega(t)$ in (\ref{phas}) 
emanates an evolution of the surfaces $\Sigma(t)$ which is
characterized by an evolution of the coordinates $z^i(\sigma^a;t)$
that embed $\Sigma(t)$ in the fluid. The general evolution
of the $z^i$ admits two contributions, there is a flow which is
normal to the surface $\Sigma(t)$ and there is a flow which is tangential
to this surface. The latter corresponds to a reparametrization
of $\Sigma(t)$ which we exclude. When we substitute the 
evolution of $z^i$ normal to the surface in (\ref{phas}) and 
assume the London limit of constant density we get
for the quantum mechanical phase
\begin{equation}
\Phi \ = \ \frac{4\pi}{3} \rho_0 Q \int\limits_{V_T}\!\!
d^3 \! \sigma \  \epsilon^{\alpha \beta \gamma}\epsilon_{ijk} 
\frac{\partial z^i}{\partial \sigma^\alpha} 
\frac{\partial z^j}{\partial \sigma^\beta}
\frac{\partial z^k}{\partial \sigma^\gamma} \ = \
\frac{4\pi}{3} Q \rho_0 V_T
\label{vol1}
\end{equation}
where $\sigma^3 = t$ and we have used the fact that the integrand 
coincides with the
volume three-form. The result should be compared to the
first term in (\ref{2d}): Clearly, (\ref{vol1}) is a 
three dimensional version of that term.

We now proceed to the general case where the periodic adiabatic 
transport ${\bf R}(s,t)$ of the closed vortex loop 
surrounds, without touching, a number of other
closed vortex loops in the fluid. This could for example relate
to the leapfrogging motion of two identical vortex loops. 
The integral (\ref{phas}) now acquires two distinct contributions.
One of these contributions is an integral that 
extends over that subregion of 
the volume $V_T$ which does not contain any 
vortices. In this subregion we are at the London limit
with $\rho({\bf x}) = \rho_0$ a constant, and the 
ensuing integral in (\ref{vol1}) leads
to a contribution where $V_T$ becomes replaced by the volume 
of the region where there are no vortices. 
The second contribution is an integral over the 
volume occupied by the vortices. In this case,
we recall that the consistency of our equations implies that 
at the location of vortices we have $\rho({\bf x}) = 0$.
Consequently the ensuing contribution to the integral
(\ref{phas}) vanishes. Combining the two contributions
we conclude
that in the presence of $N$ vortices with
circulation $n_iQ$, length $L_i$ and (average) tube
radius $\xi_i$ we get for the quantum mechanical phase
\begin{equation}
\Phi \ = \ \frac{4\pi}{3} \rho_0 Q(V_T - \pi \sum_i^N
n_i L_i \xi_i^2) 
\label{final}
\end{equation}
where the second term adds up to the total volume of the vortices
that are being encircled; the present model does not allow for a more 
detailed computation of this additional contribution as it involves Physics
at length scales which is beyond the reach of a simple Euler equation.

The final result (\ref{final}) is clearly
a direct three dimensional generalization of 
(\ref{2d}). In particular, when two identical vortex loops
leapfrog through each other, in addition to the conventional Bohm-Aharonov 
contribution to the quantum mechanical phase
we also have a contribution which is proportional to
the (average) volume of the vortices under the
leapfrogging period. It is this additional contribution
that leads to an exotic exchange statistics,
in full parallel to the two dimensional case.

The result (\ref{final}) is phenomenological
in the sense that it approximates a vortex as a circular 
tube with radius $\xi$ which supports the vorticity. 
But we argue that if we repeat our computation in a 
more involved model we arrive at the same universal 
structure. For this,
consider {\it e.g.} a complex scalar 
field $\psi$ which relates to the order parameter of 
various quantum fluids. If we write $\psi  = \sqrt{\rho} 
\exp(i\theta)$ the ensuing kinetic term acquires the 
same functional form with the kinetic term in (\ref{eulact1}),
\begin{equation}
\int d^3\!x dt \frac{i}{2} \{ \psi^* \dot \psi - \dot \psi^* \psi\}
\ = \ \int d^3\!x dt \ \rho \dot \theta
\nonumber
\end{equation}
If we repeat the present computation, the only difference
will then be that instead of a steplike density profile $\rho({\bf x})$ 
that vanishes inside the vortex tube and acquires its 
nonvanishing asymptotic London limit outside of the tube,
we now have a density profile that vanishes at the center line of the
vortex and approaches the London limit at an exponential rate
in the healing length $\xi$. Since this difference does not
introduce any qualitative changes, we conclude that
(\ref{final}) is quite universal.
    
In conclusion, we have shown that whenever the large scale
motion of closed vortex loops in a quantum fluid can
be governed by the hydrodynamical Euler equation the ensuing
statistical phase acquires an exotic contribution, which is
fully analogous to the contribution that in two dimensions
leads to anyon statistics. The simplicity of our derivation 
and the universality of the Euler equation in describing
the long distance properties of (non-relativistic) quantum fluids
is an indication that three dimensional exotic statistics 
is generic. Since vorticity is pivotal to the properties
of low viscosity and inviscid liquids and gases
we expect that our observations will have wide applicability
to the thermodynamics of fluids, from the onset of turbulence to 
dynamics of phase transitions.

\vskip 0.4cm
We thank O. Viro and L. Faddeev for discussions, and M. L\"ubcke for help
with pictures. This work has been supported by a grant VR-2003-3466
and by CDP at Uppsala University.

\end{document}